\documentstyle[aps,pre,multicol,psfig]{revtex}

\draft

\begin{document}

\title{Force fluctuations in granular materials}

\author{Alexei V\'azquez, Jos\'e Mar\'{\i}n-Antu\~na, and O.
Sotolongo-Costa}

\address{Department of Theoretical Physics, Faculty of Physics, Havana
University, Havana 10400, Cuba}

\maketitle

\begin{abstract}

Force fluctuations in granular materials are investigated. A
continuum equation is derived starting from a discrete model
proposed in the literature. The influence of boundary conditions
is investigated. For periodic boundary conditions the average
weight is found to increase linearly with depth while it
saturates to a constant value for absorbing boundary conditions,
which models the existence of walls. The scale dependencies of
the saturation weight, the saturation depth and the average
squared fluctuations are obtained. The analytical results are
compared with previous works and with numerical simulations in
one dimension.

\end{abstract}


\begin{multicols}{2}

\section{introduction}
\label{sec:introduction}

A wide variety of technical processes involve the storage and
transport of granular materials. On the other hand, granular
materials have very unusual properties which have intrigued
researches in physics over recent years \cite{general}. Of great
importance is the characterization of stress fluctuations in dry
granular materials. Optical measurements in two and three
dimensional \cite{optical} arrays of granular materials have
showed that the stress in packed granular materials is not
distributed uniformly inside the medium, but is concentrated
along "chains". Further experiments dealing with force
fluctuations in the bottom of relative small containers have
shown an exponential distribution of vertical forces (weight)
$w$ \cite{liu,mueth}.

Another interesting phenomena associated with weight
fluctuations is the problem of arching. Arching refers to the
observation that the average weight at the bottom of a column of
grains saturates at a value $W_s$ as the depth $t$ measured from
the top of the column is increased, where $W_s$ shows large
fluctuations when repeating this procedure with the same amount
of grain.  About one century ago Janssen \cite{janssen} showed,
using a simple argument, that $W(t)=W_s[1-\exp(-t/t_s)]$. $t_s$
is the saturation depth which scales as $t_s\sim L$, where $L$
is the linear horizontal size of the container.

More recently, Liu {\em et al} \cite{liu} proposed a simple
discrete model ($q$-model) which describe the essential features
of force fluctuations in granular materials. For instance,
numerical simulations of the model in $2+1$ dimensions give rise
to an exponential distribution $P(v)\sim \exp(-\lambda v)$ for
the normalized weight $v=w/t$, in agreement with experimental
observations and numerical simulations of sphere packings
\cite{liu,coppersmith}. Later Peralta-Fabi {\em et al}
\cite{pfabi} showed through numerical simulations in $1+1$
dimensions and mean field (MF) analysis that the same model, but
with different boundary conditions, explains the process of
arching, obtaining the Janssen law but with $W_s\sim L^2$ and
$t_s\sim L^2$.

Generalizations of the $q$-model has been considered
\cite{Claudin97,Nicodemi98,Socolar98,Nguyen99}. Some of them
including a local slip condition which leads to anisotropies in
the stress transmission \cite{Claudin97} and introduces
correlations \cite{Nicodemi98}. These correlations,
nevertheless, do not alter the asymptotic exponential decay of
the normalized weight distribution \cite{Nicodemi98}. On the
other hand, other authors have introduced more realistic models
which takes into account the stress fluctuations in the
horizontal cross-section \cite{Socolar98,Nguyen99}. Some of
their predictions are in agreement with those obtained for the
$q$-model, such as the exponential asymptotic decay of the
normalized weight distribution characteristic of the $q$-model.

We also count with some coarse-grained equations for the
$q$-model \cite{pfabi,Claudin98}. Claudin {\em et al}
\cite{Claudin98} obtained a diffusion equation for the weight
with a multiplicative noise and derived some analytical results
for the weight correlation function. However, their analysis was
limited to pile configurations and silos with periodic boundary
conditions. On the contrary, Peralta-Fabi {\em et al}
\cite{pfabi} the case when part of the weight is supported by
the walls of the silo, which lead to a saturation of the average
weight profile. Nevertheless, they neglected fluctuations in the
local stress transmission.

In the present work we focus our attention in obtaining a
continuum model to describe weight fluctuations and the process
of arching in granular materials, for both periodic and wall
boundary conditions. We take as starting point the $q$-model
introduced by Liu {\em et al} \cite{liu}. We obtain a
coarse-grained continuum equation of this model which is similar
to the Edwards-Wilkinson equation, where the grain weight acts
as an external force and noise is added to consider fluctuations
in shape and orientation. Boundary and initial conditions are
imposed following the definition of the discrete model. Two
particular cases are analyzed, periodic and absorbing
boundaries. In the case of periodic boundary conditions the
average weight increases linearly with depth and is independent
of the lattice size, while in the case of absorbing boundary
conditions it saturates for $t\gg t_s\sim L^2$ to $W_s\sim L^2$.
Moreover, in both cases fluctuations around the average are of
the order of $L^\zeta$, with $\zeta=(4-d)/2$.

The paper is organized as follows. In section
\ref{sec:introduction} we derive the coarse-grained equation for
the $q$-model and solve it for the case of periodic and
absorbing boundary conditions. In section \ref{sec:discussion}
our results are compared with numerical simulations in one
dimension, and with other works in the literature. Finally the
summary and conclusions are given in section \ref{sec:summary}.

\section{The model}
\label{sec:model}

\subsection{Discrete model}

Liu {\em et al} \cite{liu} introduce a simple model ($q$-model)
that describe some of the experimental observations in bead
packs.  The model assumes that the dominant physical mechanism
leading to force fluctuations is the inhomogeneity of the
packing, which causes an unequal distribution of weight on the
beads supporting a given grain. Spatial correlations in these
fractions and variations in the coordination numbers of the
grains are ignored. Only the vertical components of the forces
are considered explicitly.

Consider a $d+1$ dimensional array of grains in which each layer
has $L^d$ beads. The total weight on a given bead is transmitted
unevenly to $N$ adjacent beads in the layer underneath.
Specifically, a fraction $q_{ij}$ of the total weight supported
by the bead $j$ in layer $t$ is transmitted to bead $i$ in layer
$t+1$. Thus, a site at depth $t+1$ has weight $w_i(t+1)$, due to
its own weight $w_0$ and to weights of neighbors at depth $t$,
according to
\begin{equation}
w_i(t+1)=w_0+\sum_jq_{ij}(t)w_j(t),
\label{eq:1}
\end{equation}
where the sum runs over the $N$ neighbors, in layer $t$, of the
site $i$, in layer $t+1$. The fractions $q_{ij}(t)$ are taken to
be random variables, independent except for the constraint
\begin{equation}
\sum_i q_{ij}(t)=1,
\label{eq:2}
\end{equation}
where the sum runs over the $N$ neighbors, in layer $t+1$, of
site $j$, in layer $t$.

To complete model definition, boundary and initial conditions
must be specified. In the top layer ($l=0$) the grains have no
neighbors above and, therefore, they only support their own
weight, i.e. $w_i(0)=w_0$. On the other hand, we can take
different boundary conditions.  We consider two cases, periodic
and absorbing boundary conditions.  Absorbing boundary
conditions are more appropriate to describe the force
fluctuations in silos where part of the weight is supported
("absorbed") by the wall containing the grains. The absorbing
boundary conditions are implemented by simply imposing that
boundary sites have zero weight. If a site near the boundary
transmit part of its weight to a boundary site (i.e. to the
wall) this weight fraction is "dissipated", while the boundary
site never transmit weight to its neighbors because by
definition its weight is zero. This boundary conditions are
different but equivalent to the one used by Peralta-Fabi {\em et
al} \cite{pfabi}. On the other hand, for periodic boundary
conditions we recover the original $q$-model \cite{liu}.

\subsection{Coarse-grained equation}

To derive a coarse-grained continuum equation of eq.
(\ref{eq:1}) let us introduce the new random fraction
$Q_{ij}(t)$, such that
\begin{equation}
q_{ij}(t)=\frac{1}{N}+Q_{ij}(t),
\label{eq:3}
\end{equation}
which following (\ref{eq:2}) satisfies the constraint
\begin{equation}
\sum_i Q_{ij}(t)=0.
\label{eq:4}
\end{equation}
Substituting $q_{ij}(t)$ in eq. (\ref{eq:1}) by eq. (\ref{eq:3})
and substrating $w_i(t)$ to both sides it results that
\begin{equation}
w_i(t+1)-w_i(t)=\frac{1}{N}\sum_j[w_j(t)-w_i(t)]+w_0+\eta_i(t),
\label{eq:5}
\end{equation}
where
\begin{equation}
\eta_i(t)=\sum_j Q_{ij}(t)w_j(t).
\label{eq:6}
\end{equation}
$\eta_i(t)$ is a noise term associated with the random fractions
$Q_{ij}(t)$ and with weights at neighbor sites. In average the
contribution of $\eta_i(t)$ should be zero because of the
constraint in eq. (\ref{eq:4}).

Eq. (\ref{eq:5}) can be coarse-grained to obtain a continuum
equation for the effective $w(\vec{x},t)$. In the left hand side
we have the discrete depth derivative, while the first term in
the right hand side is the discrete Laplacian in the horizontal
direction. After coarse-graining it result that
\begin{equation}
\lambda\frac{\partial}{\partial l} w(\vec{x},t)
=\Gamma\nabla^2w(\vec{x},t)+w_0+\eta(\vec{x},t),
\label{eq:7}
\end{equation}
$\lambda$ and $\Gamma$ are coarse-grained coeficients.
$\lambda\sim a_\bot$ and $\Gamma\sim a_{||}^2/N$, where $a_\bot$
and $a_{||}$ are characteristic lengths in the vertical and
horizontal direction, respectively.

In $(1+1)$-dimensions eq. (\ref{eq:7}) is actually very similar
to the one derived in \cite{Claudin98}. Claudin {\em et al}
\cite{Claudin98} considered the explicit multiplicative nature
of the noise. However, it is not clear how their analysis can be
extended to ($2+1$)-dimensions, in particular the precise form
of the coarse-grained noise is in this case not clear for us.
Instead, we assume that $\eta(\vec{x},t)$ is a Gaussian noise
with zero mean and uncorrelated in space, with noise correlator
\begin{equation}
\langle\eta(\vec{x},t)\eta(\vec{x}^\prime,t^\prime)\rangle=
\delta(\vec{x}-\vec{x}^\prime)\Delta(t-t^\prime).
\label{eq:8}
\end{equation}
$\Delta(t)$ is a monotonically decreasing even function. We do
not know the precise form of $\Delta(t)$, but we assume it
decays to zero beyond depth $t_c$. Depending on the value of
$t_c$ we can obtain different behaviors. For $t_c\sim a_\bot$
the noise will be also uncorrelated in the vertical direction,
i.e. $\Delta(t)\sim\delta(t)$. On the contrary, if $t_c$ is much
larger than any characteristic depth then $\Delta(t)$ may be
considered constant. We expect the noise $\eta(\vec{x},t)$ to be
strongly correlated in the vertical direction. According to eq.
(\ref{eq:6}) the noise actually depends on weights at the
different sizes, which should be strongly correlated from layer
to layer. We have avoided the multiplicative nature of the noise
but in compensation we must keep the correlations in the
vertical direction. In any case we are going to consider
different choices of $\Delta(t)$ and compare the results with
numerical simulations.

\subsection{Average weight and fluctuations}

Eq. (\ref{eq:7}) can be interpreted as the equation of motion of
an interface profile $w(\vec{x},t)$, where depth plays the role
of time, under an external force $w_0$ and annealed noise
$\eta(\vec{x},t)$. This observation is very important because it
shows that force fluctuations in granular media can be described
through a more general framework, that if interface dynamics,
which has been extensively studied in the literature \cite{id}.
In this context, eq. (\ref{eq:7}) is known as the
Edwards-Wilkinson (EW) equation after \cite{EW}.

A central quantity of interest is the width $\Delta w$ of the
fluctuating "interface", given by $\Delta w^2(L,t)=\langle
L^{d}\int_0^Ld^d[w(\vec{x},t) -\langle w\rangle]^2\rangle$,
where $\langle w\rangle$ is the mean height of the interface
(mean weight). In general the "surface roughness" $\Delta w$ has
the following asymptotic behavior
\begin{equation}
\Delta w(L,t)=L^{\zeta}f(t/L^z),
\label{eq:b0}
\end{equation}
where $\zeta$ and $z$ are the roughness and dynamic exponent,
respectively. In the thermodynamic limit $L\rightarrow\infty$,
being linear, eq. (\ref{eq:7}) is readily solved via Fourier
methods. Direct integration shows that, if the noise is
spatially and temporally uncorrelated, the roughness exponent is
given by $\zeta=(2-d)/2$ \cite{EW}.

In this section we investigate the solution of eq. (\ref{eq:7})
for different boundary conditions and noise correlators
$\Delta(t)$. We are interested in the stationary solution and,
therefore, the initial condition is irrelevant.

Let first analyze the case of periodic boundary conditions. In
this case the solution can be written as
\begin{equation}
w(\vec{x},t)=\frac{w_0}{\lambda}t+y(\vec{x},t).
\label{eq:b1}
\end{equation}
After substituting this expression in eq. (\ref{eq:7}) we obtain
the following equation for $y(\vec{x},t)$
\begin{equation}
\lambda\partial_t y(\vec{x},t)
=\Gamma\nabla^2y(\vec{x},t)+\eta(\vec{x},t).
\label{eq:b2}
\end{equation}
The solution of this equation clearly has zero average, i.e.
$\langle y(\vec{x},t)\rangle=0$, and therefore
\begin{equation}
\langle w(\vec{x},t)\rangle=\frac{w_0}{\lambda}t.
\label{eq:b3}
\end{equation}
The average weight thus increases linearly with $t$ and no
saturation is observed.

On the contrary, for absorbing boundary conditions the solution
cannot be proposed as in eq. (\ref{eq:b1}). In this case weight
dissipation at the boundary leads to a stationary state where
the average weight saturates. In the language of interface
dynamics this situation correspond with an elastic interface
pinned at the boundary under an external force $w_0$. In this
case is better to look for a solution of the form
\begin{equation}
w(\vec{x},t)=W(\vec{x})+y(\vec{x},t),
\label{eq:b4}
\end{equation}
where $W(\vec{x})$ is the solution of the stationary problem
\begin{equation}
\Gamma\nabla^2W(\vec{x})+w_0=0,
\label{eq:b5}
\end{equation}
with homogeneous boundary conditions ($W|_{\text{boundary}}=0$),
and $y(\vec{x},t)$ satisfies eq. (\ref{eq:b2}) but with
homogeneous boundary conditions.

Again, $y(\vec{x},t)$ has zero average and, therefore, the mean
weight in the stationary state is $W(\vec{x})$. The solution of
eq. (\ref{eq:b5}) can be easily obtained for certain geometries.
In $d=1$ we can look for the solution in the interval $0<x<L$
with $W(0)=W(L)=0$, obtaining
\begin{equation}
W(x)=\frac{w_0L^2}{2\Gamma}\left(\frac{x}{L}-
\frac{x^2}{L^2}\right).
\label{eq:b6}
\end{equation}
For $d=2$, $d=3$ and $d>3$ (only the case $d=2$ has a physical
realization) we can look for the solution in a circle, sphere
and hyper-sphere of radius $L$ such that $W=W(r)$, with $0<r<L$
and the boundary condition $W(L)=0$, where $r$ is the distance
to the center. The solution is in this case given by
\begin{equation}
W(r)=\frac{w_0L^2}{2d\Gamma}\left(1-\frac{r^2}{L^2}\right).
\label{eq:b7}
\end{equation}
Thus, in any dimension, $L$ is the characteristic length of the
system and the saturation weight scales as $L^2$.

Now we are going to analyze the fluctuations around the average,
described by eq. (\ref{eq:b2}) with the corresponding boundary
conditions and noise correlator. A formal solution of eq.
(\ref{eq:b2}) is given by
\begin{equation}
y(\vec{x},t)=\int_0^t dt^\prime\int d^dx^\prime 
G_d(\vec{x},\vec{x^\prime},t-t^\prime;L)\eta(\vec{x}^\prime,t^\prime),
\label{eq:b8}
\end{equation}
where $G_d(\vec{x},\vec{x^\prime},t;L)$ is the Green function of
corresponding boundary problem, which depends on the dimension
of the horizontal space $d$ and on the linear size of the system
$L$. The precise form of the Green function can be obtained only
for some suitable geometries, however, in general it satisfies
the scaling relation
\begin{equation}
G_d(\vec{x},\vec{x^\prime},l;L)=\frac{1}{L^d}
g_d\left(\frac{\vec{x}}{L},\frac{\vec{x}^\prime}{L},
\frac{t}{L^2}\right),
\label{eq:b9}
\end{equation}

\begin{figure}\narrowtext
\centerline{\psfig{file=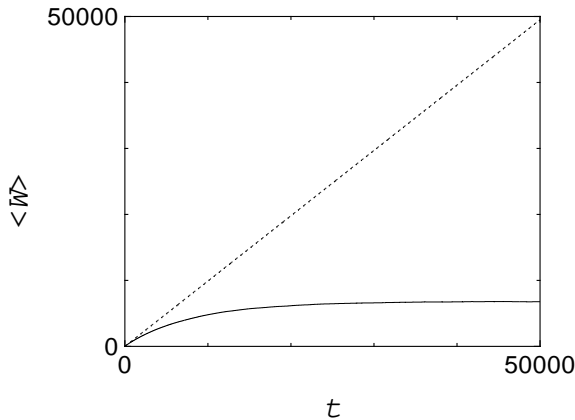,width=3in}}
\caption{Global average weight as a function of depth obtained
from numerical simulations in one dimension. The continuous
(dashed) line corresponds with absorbing (periodic) boundary
conditions. In the case of absorbing boundary conditions the
average weight saturates after a characteristic depth $t_s$
while it increases linearly with depth for periodic boundary
conditions.}
\label{fig:1}
\end{figure}

\noindent
where $g_d(\vec{x},\vec{x}^\prime,t)$ depends on the spatial
dimension and boundary conditions and $L$ is a characteristic
linear dimension of the system. $g_d$ can be obtained exactly,
for instance, in one dimension and in a systems with radial
symmetry.

Using equations (\ref{eq:b8}), (\ref{eq:b9}), and (\ref{eq:8})
we compute the "surface roughness" $\Delta w$, obtaining the
scaling relation in eq. (\ref{eq:b0}) with $z=2$ and a roughness
exponent $\zeta$ which depends on the choice of $\Delta(t)$. If
the noise is uncorrelated in the vertical direction $t$
($t_c\rightarrow0$) then
\begin{equation}
\zeta=\frac{2-d}{2},
\label{eq:b10}
\end{equation}
which corresponds with the EW universality class. On the
contrary, if it is strongly correlated in the vertical direction
($t_c\rightarrow\infty$) then
\begin{equation}
\zeta=\frac{4-d}{2}.
\label{eq:b11}
\end{equation}
In both cases the exponent $\zeta$ is independent of the choice
of boundary conditions, and they are identical to the exponents
obtained from the Fourier analysis for a $L\rightarrow\infty$
system. Thus, while the average profile is strongly dependent on
the boundary conditions, the "surface roughness" only depends on
the noise correlator.

\section{Numerical simulations and discussion}
\label{sec:discussion}

To test results obtained in the previous section we have
performed numerical simulations of the discrete model in one
dimension ($N=2$) with a uniform $q$- distribution. We have used
$w_0=1$ and lattice sizes $L=50$, $100$, $200$ and $400$.
Average were taken over 1000 realizations for the three smallest
lattice sizes and over 100 realizations 

\begin{figure}\narrowtext
\centerline{\psfig{file=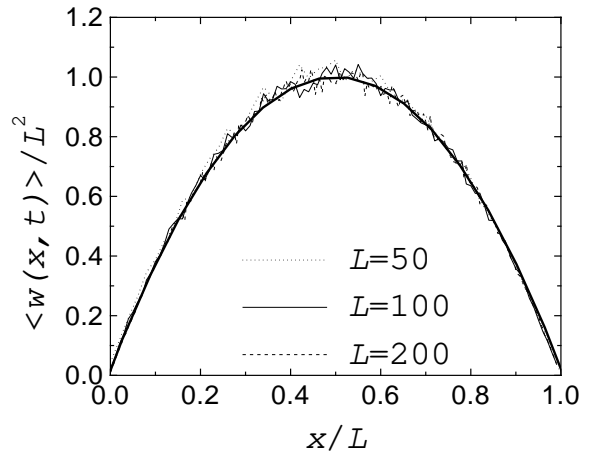,width=3in}}
\caption{Weight profile in the stationary state for absorbing
boundary conditions. The choice of scaled variables is explained
in the text. The heavy line is a fit to the quadratic dependency
$P(x)=a+bx-cx^2$, with $a=0.02\pm0.01$, $b=3.91\pm0.1$ and
$c=3.92\pm0.1$.}
\label{fig:2}
\end{figure}

\noindent
for the largest one.
Different magnitudes associated with force fluctuations were
computed.

Our analytical approach reveals that the existence of absorbing
boundaries  is necessary to obtain a saturation weight,
otherwise the weight increases linearly with depth. In fig.
\ref{fig:1} we have plotted the global average weight
\begin{equation}
W(t)=\left\langle\frac{1}{L}\sum_{i=1}^Lw_i(t)\right\rangle
\label{eq:c1}
\end{equation}
at depth $t$, for periodic and absorbing boundary conditions.
The agreement with our prediction becomes evident.

Moreover, in the case of absorbing boundary conditions, we have
computed the average weight profile after saturation which in
one dimension is given by eq. (\ref{eq:b6}). The average weight
after saturation as a function of lattice position $x$ is shown
in fig. \ref{fig:2}, for different lattice sizes. The scaled
variables $w/L^2$ vs. $x/L$ have been used as suggested by eq.
(\ref{eq:b6}). All the curves collapse in a single plot showing
that our choice of scaled variables is correct. Moreover, the
scaled plot was fitted to the quadratic dependency
$P(x)=a+bx-cx^2$, with $a=0.02\pm0.01$, $b=3.91\pm0.1$ and
$c=3.92\pm0.1$. These fitting parameters are in very good
agreement with eq. (\ref{eq:b6}) since they satisfy $a\ll b$ and
$b\approx c$. We have thus obtained the correct profile for the
saturation weight.

\begin{figure}\narrowtext
\centerline{\psfig{file=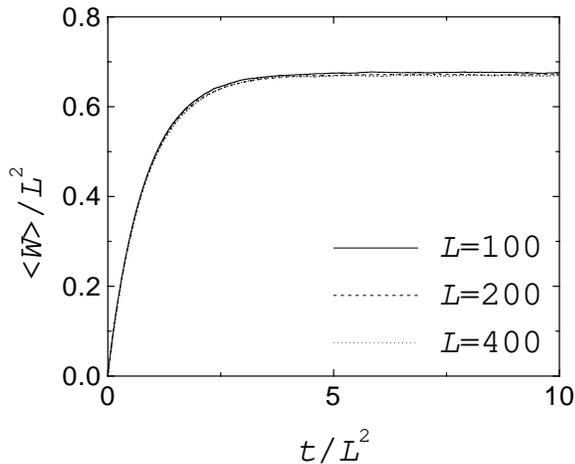,width=3in}}
\caption{Average weight as a function of depth for different
lattice sizes. The choice of scaled variables is explained in
the text.}
\label{fig:3}
\end{figure}

To investigate the transient region before saturation we have
computed the global average weight as a function of depth. The
result for different lattice sizes is shown in fig. \ref{fig:3}.
The scaled variables $W(t)/L^2$ vs.$t/L^2$ have been used. For
small depths $W(t)$ scales linearly with $t$, as for periodic
boundary conditions. In this region the force chains starting
from bulk sites have not reached the boundary and, therefore,
the weight dissipation at the boundary 

\noindent
is negligible. This
transient behavior was previously observed in numerical
simulations by Peralta-Fabi {\em et al} in one dimension
\cite{pfabi}.

\begin{figure}\narrowtext
\centerline{\psfig{file=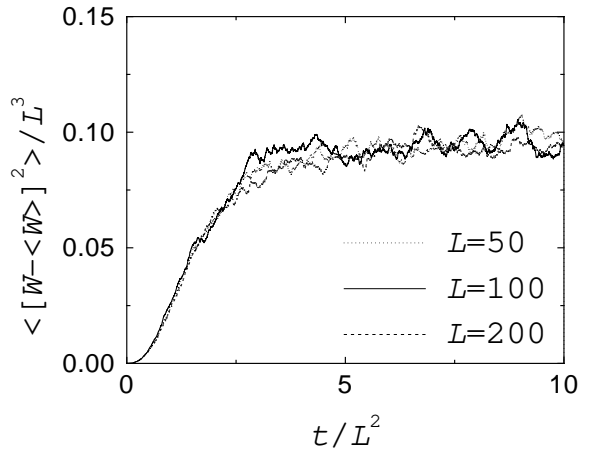,width=3in}}
\caption{Average squared fluctuations as a function of depth for
different lattice sizes. The choice of scaled variables is
explained in the text.}
\label{fig:4}
\end{figure}

The average squared fluctuations (interface roughness) $\Delta
W^2=\langle(W-\langle W\rangle)^2\rangle$ as a function of $t$
are shown in fig. \ref{fig:4}. Using the scaled variables
$\Delta W^2/L^{2\zeta}$, with $\zeta=3/2$, vs. $t/L^2$ we obtain
a good data collapse. This scaled plot is very sensitive to the
choice of $\zeta$, a good data collapse is only obtained for
$\zeta=1.50\pm0.05$. If the noise is uncorrelated in the
vertical direction then $\zeta=1/2$, which is much smaller than
our numerical estimate. On the contrary, if the noise is
strongly correlated in the vertical direction then $\zeta=3/2$,
in very good agreement with our numerical estimate. When we
avoid the multiplicative character of the noise we guess that,
in compensation, strong correlations in the vertical direction
should be considered. This supposition is now supported by
numerical simulations.

The simplicity of the continuum model we have proposed in the
previous section allowed us to obtain some analytical results.
We have avoided the complicated form of the noise, associated
with the fluctuations in the fractions $Q_{ij}$, introducing an
"annealed" noise with correlations in the vertical direction.
Within this approximation we have obtained the average weight in
the stationary state and characterized the fluctuations around
the average, for both periodic and absorbing boundary
conditions.

In earlier numerical simulations by Liu {\em et al} \cite{liu},
they considered periodic boundary conditions and compute the
distribution of normalized weight $v=w/t$. They observe that the
distribution of normalized weight $P(v)$ was independent of $t$
for large $t$. This numerical result was later corroborated by
Coppersmith {\em et al} using a MF theory for the discrete model
\cite{coppersmith}.  According to our approach, for periodic
boundary conditions the average weight is given by $w_0
t/\lambda$ and, therefore, $\langle v\rangle=w_0/\lambda$. Since
the average normalized weight is independent of $t$ so should be
its distribution, in agreement with the numerical sim$-$

\noindent
ulations and MF theory of the discrete model.

The robustness of the asymptotic exponential decay of the
normalized weight distribution $P(v)$ is one of the main
features of the $q$-model. We thus check if this behavior is
still observed when absorbing boundary conditions are
considered. Before going to the numerical results we should
remember that in this case for $t\gg t_s$ the average weight
does not increases linearly with $t$ but scales as $L^2$. Hence,
the appropriate choice of normalized weight is in this case
$v=w/L^2$.

\begin{figure}\narrowtext
\centerline{\psfig{file=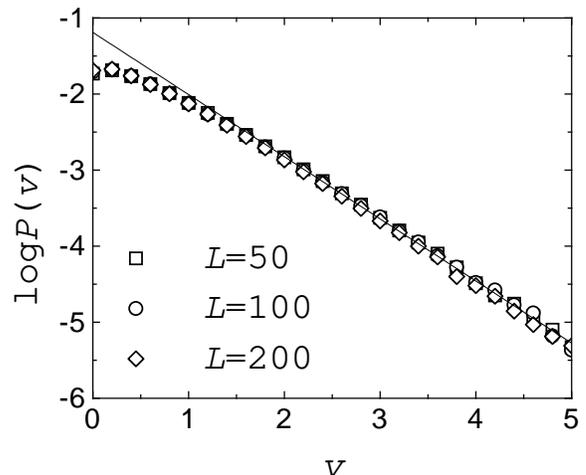,width=3in}}
\caption{Distribution of normalized weight $v=w/L^2$, after
saturation $t\gg t_s$, for the $q$-model with absorbing boundary
conditions. The straight line is a fit to an exponential decay.}
\label{fig:5}
\end{figure}

With this remark, in fig. (\ref{fig:5}) we show the distribution
of normalized weight for different lattice sizes. First we
notice that $P(v)$ is independent of lattice size, supporting
our choice of normalized weight. Moreover, the resulting
universal curve clearly show an exponential decay for large $v$.
However, for small $v$ the data does not display the linear
behavior predicted by the $q$-model for a uniform distribution
\cite{liu,coppersmith}. This discrepancy for small $v$ is just a
consequence of the different boundary conditions.

On the other hand, in the case of absorbing boundary conditions
our analytical approach is in agreement with previous MF theory
by Peralta-Fabi {\em et al} \cite{pfabi} in one dimension.
Peralta-Fabi {\em et al} analyzed the case $Q_{ij}\equiv0$,
obtaining the scaling dependencies $W_s\sim L^2$ and $t_s\sim
L^2$. We have shown that these scaling relations are valid in
larger dimensions and are independent of the noise. However,
within the MF theory by Peralta-Fabi {\em et al} one cannot
determine the scale dependency of the average squared
fluctuations $\Delta w^2$, which constitutes one of our main 
results. 

The scaling dependencies $W_s\sim L^2$ and $t_s\sim L^2$ are,
nevertheless, in contradiction with the linear scaling obtained
in classical theories \cite{janssen}, some generalizations of
the $q$-model \cite{Socolar98} and even in experimental
observations \cite{Clement97}. However, this discrepancy cannot
be attributed to our continuum analysis, which shows a very good
agreement with the numerical simulations of the $q$-model.

\section{Summary and conclusions}
\label{sec:summary}

We have obtained a coarse-grained equation of the discrete model
introduced by Liu {\em et al} to describe force fluctuations in
granular media. The multiplicative nature of the noise has been
considered assuming strong correlations in the vertical
direction. The stationary and transient behavior were obtained
analytically for periodic and absorbing boundary conditions.

In this way we have demonstrated that the existence of walls,
modeled by the absorbing boundary conditions, are a necessary
condition to obtain an stationary state were the average weight
is independent of depth. We have also shown that the the scaling
of the saturation weight $W_s\sim L^2$ and depth $t_s\sim L^2$
with lattice size are valid in any dimension and are independent
of the noise, generalizing previous MF calculations in one
dimension.

For the first time we have obtained the scale dependency of the
average squared fluctuations. The comparison of this scale
dependence with the one obtained in the numerical simulations
support our guess about the existence of strong correlations of
the noise in the vertical direction.

\vskip 0.5in

\end{multicols}

\end{document}